\documentclass[12pt,a4paper]{article}

\usepackage{amsmath}
\usepackage{amsfonts}
\usepackage{amssymb}
\usepackage{amsthm}
\usepackage{slashed} 
\usepackage{graphicx}
\usepackage{epstopdf}
\usepackage[dvipsnames,svgnames,x11names,hyperref]{xcolor}
\usepackage[pdftex=true,colorlinks=false]{hyperref}
\usepackage[figure,table]{hypcap} 
\hypersetup{
  	citebordercolor={red},
    linkbordercolor={blue},
    urlbordercolor={ForestGreen},
}

\newcommand{\ISL}{\mathrm{ISL}}
\newcommand{\isl}{\mathrm{isl}}

\newcommand{\SL}{\mathrm{SL}}
\newcommand{\sll}{\mathrm{sl}}

\newcommand{\SO}{\mathrm{SO}}
\newcommand{\ISO}{\mathrm{ISO}}

\newcommand{\so}{\mathrm{so}}

\newcommand{\dd}{\mathrm{d}}      
\newcommand{\D}{\mathrm{D}}

\newcommand{\R}{\mathbb{R}}   
\newcommand{\C}{\mathbb{C}}

\newcommand{\diag}{\mathrm{diag}}

\numberwithin{equation}{section}

\theoremstyle{definition}

\begin{document}

\title{Hamiltonian analysis of self-dual gauge gravity}

\author{Steven Kerr \\
Perimeter Institute for Theoretical Physics\\
31 Caroline Street North\\
Waterloo, Ontario, N2J 2Y5, Canada\\
\\
skerr@perimeterinstitute.ca
}

\date{}
\maketitle

\begin{abstract}
The Hamiltonian analysis of the self-dual gauge gravity theory is carried out. The resulting canonical structure is equivalent to that of self-dual gravity.
\end{abstract}

\newpage

\section{Introduction}
Canonical quantisation of general relativity is a long-standing problem. The ADM formalism \cite{ADM} is a canonical analysis of general relativity using metric variables. The resulting form of the Hamiltonian constraint is a complicated, non-polynomial expression in the canonical variables. As a result, attempts at constructing the corresponding quantum operator quickly run into problems. In the 80's, Ashtekar reformulated gravity in terms of new variables that significantly simplify the form of the constraints \cite{Ashtekar}. This had led to the `loop quantum gravity' research programme.

In \cite{Witten}, Witten quantised $(2+1)$-dimensional gravity exploiting the fact that the first order action is equivalent to a Chern-Simons gauge theory with gauge group $\ISO(2,1)$ for zero cosmological constant, and $\SO(3,1)/\SO(3,2)$ for positive/negative cosmological constant. This discovery sparked interest in whether a similar strategy might work in $3+1$ dimensions. The first order formalism for $(3+1)$-dimensional gravity was first explicitly formulated as a gauge theory with gauge group $\SO(4,1)/\SO(3,2)$ by Macdowell and Mansouri \cite{MM}. However, canonical analysis of this theory reveals second-class constraints which are difficult to deal with when quantising \cite{MM hamiltonian}. A closely related but simpler gauge theory was discovered by Pagels \cite{Pagels} in the $\SO(4,1)/\SO(3,2)$ case, and generalised to $\ISO(3,1)$ by Grignani and Nardelli \cite{GN}. The purpose of this paper is to carry out the Hamiltonian analysis for the self-dual variation of this theory. It is shown that the resulting canonical structure is identical to that of self-dual gravity.

\section{Conventions, notation and identities}

Upper calligraphic indices $\mathcal{A},\mathcal{B} \ldots = 0 \ldots 4 $ transform under a five-dimensional vector representation of $\ISO(3,1)$ as $x^{'\mathcal{A}} = M^{\mathcal{A}}_{\phantom{\mathcal{A}}\mathcal{B}} x^{\mathcal{B}}$, with 

\begin{align}
M^{\mathcal{A}}_{\phantom{\mathcal{A}}\mathcal{B}} =  \left( \begin{array}{cc} M^A_{\phantom{A} B} & t^A \\
0 & 1 \\
\end{array} \right). \label{vector rep}
\end{align}
Here $M^A_{\phantom{A} B} \in \SO(3,1)$, $t^A \in \R^4$ is the translation, and capital indices $A,B \ldots = 0 \ldots 3$ are the restriction of $\mathcal{A},\mathcal{B} \ldots$ to run from $0$ to $3$. Taking the particular value $x^4=0$ gives a subrepresentation in which the translations act trivially.

Lower calligraphic indices transform under the covector representation of $\ISO(3,1)$ as $y'_{\mathcal{A}} = P_{\mathcal{A}}^{\phantom{\mathcal{A}}\mathcal{B}} y_{\mathcal{B}}$, with 

\begin{align}
P_{\mathcal{A}}^{\phantom{\mathcal{A}}\mathcal{B}} =  \left( \begin{array}{cc} P_A^{\phantom{A} B} & 0 \\
-s^A  {P}^{\phantom{A} B}_{A} & 1 \\
\end{array} \right). \label{covector rep}
\end{align}
Here $P_A^{\phantom{A} B} \in \SO(3,1)$ and $s^A \in \R^4$ is the translation.

The invariant bilinear form that can be used to contract two $\ISO(3,1)$ vectors, whose transformation law is given by \eqref{vector rep}, is  $\eta_{\mathcal{A} \mathcal{B}} = \diag(0,0,0,0,1)$. The invariant bilinear form that can be used to contract two $\ISO(3,1)$ covectors, whose transformation law is given by \eqref{covector rep}, is $\eta^{\mathcal{A} \mathcal{B}} = \diag(-1,1,1,1,0)$.

 Lower case indices $a, b \ldots = 1\ldots 3$ are the restriction of $A,B \ldots = 0 \ldots 3$ to run from $1$ to $3$. 

Spacetime indices are denoted by Greek letters, $\alpha, \beta \ldots = 0 \ldots 3$. Space indices $i,j \ldots = 1 \ldots 3$ are the restriction of spacetime indices to run from $1$ to $3$.

There is a potential ambiguity when objects with multiple indices are written in components. A subscript $0$ for example could indicate an $\ISO(3,1)$ index or a spacetime index taking the value $0$. However, it will always be clear from context which type of index has been fixed.

The spacetime metric signature is $\diag(-1,1,1,1)$. The spacetime Levi-Civita symbols $\epsilon^{\alpha \beta \gamma \delta }$, $\epsilon_{\alpha \beta \gamma \delta }$ are completely antisymmetric and defined so that $\epsilon^{0123} = \epsilon_{0123} = 1$ in all coordinate systems.

The $\ISO(3,1)$ Levi-Civita symbols $\epsilon^{\mathcal{A} \mathcal{B} \mathcal{C} \mathcal{D} \mathcal{E}}$, $\epsilon_{\mathcal{A} \mathcal{B} \mathcal{C} \mathcal{D} \mathcal{E}}$ are completely antisymmetric and defined so that $\epsilon^{01234} = - \epsilon_{01234} = 1$. These are invariant tensors since the transformations \eqref{vector rep}, \eqref{covector rep} both have determinant one. We define $\epsilon^{ABCD} = \epsilon^{\mathcal{A} \mathcal{B} \mathcal{C} \mathcal{D} 4}$, $\epsilon_{ABCD} = \epsilon_{\mathcal{A} \mathcal{B} \mathcal{C} \mathcal{D} 4} $, so that $\epsilon^{0123} = - \epsilon_{0123} = 1$. Finally we define $\epsilon^{abc} = \epsilon^{0abc}$, $\epsilon_{abc} = -\epsilon_{0abc}$, so that $\epsilon^{123} = \epsilon_{123} = 1$.

The complexification of a real Lie algebra is obtained informally by `taking the underlying field to be $\C$'. For $\ISO(3,1) = \ISO(3,1,\R)$, the corresponding complex Lie group is denoted $\ISO(3,1, \C)$. The representation theory constructed above carries over straightforwardly to the complex case.

The dual of an object $K^{AB}$ with two capital Latin indices is defined by

\begin{align}
{K^*}^{AB} = \frac{1}{2} \epsilon^{ABCD} K_{CD}.
\end{align}
Assuming $K^{AB} = -K^{BA}$ is antisymmetric, it may be written as a sum of `self-dual' and `antiself-dual' parts,

\begin{align}
K^{AB} = K^{+AB} + K^{-AB},
\end{align}
where $K^{\pm AB} = \frac{1}{2}\left( K^{ AB} \mp i{K^*}^{AB}\right)$, satisfying

\begin{align}
(K^+)^* = iK^+, \quad (K^-)^* = -iK^-. \label{dual}
\end{align}
In the $\SO(4)$ case with an `all positive' bilinear form, there are no factors of $i$ in \eqref{dual}, and this is the origin of the `self-dual' and `antiself-dual' terminology.

We define $K^a := K^{a0} = -K^a_0 $. There exist relations between the different components of $K^{+AB}$,

\begin{align}
K^{+ ab} = -i \epsilon^{abc} K^+_c. \label{dual relations}
\end{align} 

The four-dimensional spacetime manifold $M$ will be assumed to admit a foliation of the form $M = \Sigma \times \R$, with $\Sigma$ a $3$-manifold without boundary. The metric may be written in block components as follows

\begin{align}
g_{\alpha \beta} = \left( \begin{array}{cc} -N^2 + N_i N^i & N_i \\
N_i & g_{ij} \end{array} \right), \quad
g^{\alpha \beta} = \left( \begin{array}{cc} -\frac{1}{N^2} & \frac{N^i}{N^2} \\
\frac{N^i}{N^2} & g^{ij} \end{array} \right).
\end{align}
Here $N$, $N^i$ are called the lapse and shift functions respectively. We will use the `triangular gauge', in which the components of the frame field are as follows

\begin{align}
e_{\alpha A} = \left( \begin{array}{cc} N & N_i e^i_a \\
0 & e_{ia} \end{array} \right), \quad
e^{\alpha A} = \left( \begin{array}{cc} \frac{1}{N} & 0 \\
-\frac{N^i}{N} & e^{ia} \end{array} \right). \label{triangular gauge}
\end{align}

The complexification of the Lie algebra $\so(3,1)$ is $\so(3,1, \C) = \sll(2, \C) \oplus \sll(2, \C) $. The $\sll(2, \C)$ factors are given by the self-dual and antiself-dual parts of the $\so(3,1,\C)$ connection. The $\so(3,1, \C)$ curvature tensor is defined by 

\begin{align}
R^{A}_{jk B} = \frac{1}{2} \left( \partial_j \omega_{kB}^{A} - \partial_k \omega_{jB}^{A} + \omega^{A}_{jC} \omega^C_{kB} - \omega^{A}_{kC} \omega^C_{jB} \right),
\end{align}
with $\omega^A_{k B}$ the complexified spin connection.

It can be shown that the self-dual part of the $\so(3,1, \C)$ curvature tensor is equal to the curvature of the self-dual part of the $\so(3,1, \C)$ connection,

\begin{align}
R^{+AB}_{jk} = \frac{1}{2} \left( \partial_j \omega_k^{+AB} - \partial_k \omega_j^{+AB} + \omega^{+A}_{j\phantom{A}C} \omega^{+CB}_{k} - \omega^{+A}_{k\phantom{A}C} \omega^{+CB}_{j} \right).
\end{align}
The self-dual part of the complexified spin connection is an $\sll(2, \C)$ connection, and

\begin{align}
R_{jk}^{+a} = \frac{1}{2} \left( \partial_j \omega_k^{+a} -  \partial_k \omega_j^{+a} + 2i \epsilon^{abc} \omega^+_{jb} \omega^+_{kc} \right).
\end{align}

\section{Hamiltonian analysis}

The action of self-dual gauge gravity is

\begin{align}
S = \int_M (\D \phi)^{\mathcal{A}} 
\wedge  (\D \phi)^{\mathcal{B}} 
\wedge F^{\mathcal{C}\mathcal{D}} 
  \epsilon_{\mathcal{A} \mathcal{B} \mathcal{C} \mathcal{D} \mathcal{E}} \phi^{\mathcal{E}}  . 
 \label{action}
\end{align}
The gauge group is $\ISL(2, \C) = \SL(2, \C) \ltimes \C^4 $, the semidirect product of the complex translation group $\C^4$ with the $\SL(2, \C)$ group generated by the self-dual part of an $\so(3,1,\C)$ connection. The action of $\SL(2, \C)$ on $\C^4$ in the semidirect product is given by the four-dimensional representation of $\SL(2, \C)$ inherited from the defining four-dimensional representation of $\SO(3,1,\C)$. $\phi^{\mathcal{A}}$ is a multiplet of complex scalar fields in the vector representation \eqref{vector rep} of $\ISL(2, \C)$, with the gauge invariant constraint $\sqrt{\phi^{\mathcal{A}} \phi^{\mathcal{B}} \eta_{\mathcal{A} \mathcal{B}}} = \phi^4=1$ imposed (the square root is the positive square root). The $\isl(2, \C)$ connection in block form is $A^{\mathcal{A}}_{\phantom{\mathcal{A}}\mathcal{B}}=\left( \begin{array}{cc}   A^A_{\phantom{A} B} & E^A\\
0 & 0 \end{array} \right)$, with $A^A_{\phantom{A} B}$ a self-dual connection generating $\SL(2, \C)$, and $E^A$ generating $\C^4$. The covariant derivative is

\begin{align}
 (\D \phi)^{\mathcal{A}} = \dd \phi^{\mathcal{A}} + A^{\mathcal{A}}_{\phantom{\mathcal{A}}\mathcal{B}} \phi^{\mathcal{B}}   =  \left( \begin{array}{c} \dd \phi^A + A^A_{\phantom{A} B} \phi^B + E^A \\
0 \end{array} \right).
\end{align} 
The curvature is

\begin{align}
F^{\mathcal{C}}_{\phantom{\mathcal{C}} \mathcal{D}} = \left( \begin{array}{cc} \dd A^C_{\phantom{C} D} + A^C_{\phantom{C} B} \wedge A^B_{\phantom{B} D} & \dd E^C + A^{C}_{{\phantom{C}}B} \wedge E^B  \\
0 & 0  \end{array} \right).
\end{align}
Raising the second index using the metric $\eta^{\mathcal{A} \mathcal{B}}$ gives
\begin{align}
F^{\mathcal{C} \mathcal{D}} = F^{\mathcal{C}}_{\phantom{\mathcal{C}} B} \eta^{\mathcal{B} \mathcal{D}} = \left( \begin{array}{cc} \dd A^{CD} + A^{C}_{\phantom{C} B} \wedge A^{BD}  & 0  \\
0 & 0  \end{array} \right).
\end{align}

By a translational gauge transformation, we may take 

\begin{align}
\phi^{\mathcal{A}} \rightarrow \left( \begin{array}{c} 0 \\
0 \\
0 \\
 0 \\
 1 \\
 \end{array} \right). \label{physical gauge}
\end{align} 
This is called `physical gauge'. In this gauge, $A^{A}_{\phantom{A} B}$ is identified with $\omega^{+A}_{\phantom{+A} B}$, the self-dual part of the complex spin connection, and $E^A$ is identified with the complexified frame field $e^A$. The action is

\begin{align}
S = \int_M e^A \wedge e^B \wedge R^{+CD}  \epsilon_{ABCD},
\end{align}
with $R^{+CD} = \dd \omega^{+CD} + \omega^{+C}_{\phantom{+C} B} \wedge \omega^{+BD} $. This is the self-dual Holst action with zero cosmological constant.

One only obtains non-zero contributions to the action \eqref{action} when the index $\mathcal{E}=4$, so

\begin{align}
S = \int_M  (\D \phi)^{A} 
\wedge  (\D \phi)^{B} 
\wedge {F}^{CD} 
   \epsilon_{ABCD}  . 
 \label{action1}
\end{align}
The first step in passing over to the Hamiltonian formulation is to separate the action into variables whose time derivative appears, and variables whose time derivative does not appear,

\begin{align}
S &= \int_M \dd^3 x \hspace{0.6pt} \dd t \bigg{\{}   2\dot{\phi}^A  (D_{i} \phi)^B F^{CD}_{jk } \epsilon_{ABCD} \epsilon^{ijk} + \dot{A}^{CD}_i (D_{j} \phi)^A (D_{k} \phi)^B  \epsilon_{ABCD} \epsilon^{ijk}  \nonumber \\ 
  &+  2E_0^A (D_{i} \phi)^B F^{CD}_{jk }\epsilon_{ABCD} \epsilon^{ijk} + A^{AB}_0 \Big[  2\phi_B  (D_{i}\phi)^C F^{DE}_{jk }  \epsilon_{ACDE} \epsilon^{ijk}   \nonumber \\
  &+ (D_{i}\phi)^C (D_{j}\phi)^D A^{\phantom{kB} E}_{k B} \epsilon_{CDAE} \epsilon^{ijk} - (D_{i}\phi)^C (D_{j}\phi)^D A^E_{k A} \epsilon_{CDEB} \epsilon^{ijk} \nonumber \\ 
& + 2( \partial_{i} ( D_{j} \phi)^C) (D_{k}\phi)^D  \epsilon_{CDAB} \epsilon^{ijk} \Big]  \bigg{\}}. \label{separation}
\end{align}
Upon using the relations \eqref{dual relations} that link the different components of the connection $A^A_{\phantom{A} B}$ and substituting in the canonical momenta, this becomes

\begin{align}
S = \int_M \dd^3 x \hspace{0.6pt} \dd t \left\{ \dot{\phi}^A \pi_A + \dot{A}_i^d p^i_d + E_0^A \pi_A + A_0^d \left[ \pi_d \phi_0 - \phi_d \pi_0 -i \epsilon_{dab} \pi^a \phi^b + D_i p^i_d \right] \right\},
\end{align}
with $D_i p^i_d = \partial_i p^i_d +2i \epsilon_{dab} A^a_i p^{ib}  $, and

\begin{align}
\pi_A &= \frac{\partial \mathcal{L}}{\partial \dot{\phi}^A} = 2 (D_{i}\phi)^B F^{CD}_{jk}  \epsilon_{ABCD} \epsilon^{ijk}, \label{pi constraint} \\
p^i_d &=  \frac{\partial \mathcal{L}}{\partial \dot{A^d_i}} =  2(D_j\phi)^a (D_k\phi)^b \epsilon_{abd} \epsilon^{ijk} + 4 i  (D_j\phi)^0 (D_k\phi)_d \epsilon^{ijk}. \label{p definition}
\end{align}
The first equation \eqref{pi constraint} defines a primary constraint. We note the equality $(\D \phi)^A= e^A$, which is due to the fact that $(\D \phi)^A$ is invariant under translations. Using the triangular gauge \eqref{triangular gauge} to evaluate $p^i_d$ gives

\begin{align}
p^i_d = 4 (D^i \phi)_d  \det(D_i \phi_a) =  4 e^i_d \det(e_{ia}). \label{p e relation}
\end{align}
Thus $p^i_d$ is determined fully by $e^i_d$, and \eqref{p definition} does not define a primary constraint.

The Hamiltonian is

\begin{align}
H = \int_{\Sigma} \dd^3 x \left( - E_0^A \pi_A - A_0^d \sigma_d \right),
\end{align}
where $\sigma_d := \pi_d \phi_0 - \phi_d \pi_0 -i \epsilon_{dab} \pi^a \phi^b + D_i p^i_d$. The constraints are

\begin{align}
\pi_A &= 0, \nonumber \\
c_A &:= \pi_A -  2 (D_{i}\phi)^B F^{CD}_{jk} \epsilon_{ABCD} \epsilon^{ijk} = 0, \nonumber \\
\sigma_d&=0.
\end{align}
Using \eqref{p e relation}, an equivalent set of constraints is

\begin{align}
\pi_A &= 0,\nonumber \\
C_a &:=  p_i^b F_{jk}^c \epsilon^{ijk} \epsilon_{abc} = 0 ,\nonumber \\
C &:= p_i^b F_{jk b} \epsilon^{ijk} = 0 \nonumber \\
G_d&:= D_i p^i_d = 0,
\end{align}
where $p_i^b$ is the inverse momentum satisfying $p_i^b p_c^i = \delta^b_c$. These constraints are related to the diffeomorphism and Hamiltonian constraints of self-dual gravity by

\begin{align}
C_a p^a_j &= 2p p^i_b F_{ij}^b =  2p \mathcal{H}_j, \\
 C &= p p^i_a p^j_b F_{ij}^c \epsilon^{\phantom{c} ab}_c = p\mathcal{H},
\end{align}
where $p = \det(p_{ia})$, $\mathcal{H}_j = p_a^i F_{ij}^a$ and $\mathcal{H} = p^i_a p^j_b F_{ij}^c \epsilon^{\phantom{c} ab}_c$. Thus the constraint algebra closes, and the canonical structure is essentially identical to that of self-dual gravity.

$\sigma_d$ generates $\SL(2, \C)$ transformations, while $\pi_A$ generates translations,

\begin{align}
\{ \sigma_d[\lambda^d], A_i^b(x) \} &= D_i \lambda^b(x), \\ 
\{ \sigma_d[\lambda^d], p^i_b(x) \} &= 2i \epsilon_{bad} p^{ia} (x) \lambda^d(x) ,\\
\{ \pi_A[E_0^A], A_i^b(x) \} &= \{ \pi_A[E_0^A], p_i^b(x) \} = 0, \\
\{ \pi_A[E_0^A], \phi^B(x) \} &= - E_0^B(x).
\end{align}
$\sigma_d$ and $\pi_A$ have the following Poisson brackets

\begin{align}
\{ \pi_A[E_0^A], \pi_A[F_0^A] \} &= 0, \\
\{ \sigma_d[\lambda^d], \pi_A[E_0^A] \} &= \int_{\Sigma} \dd^3 z \left( -\eta_{A0} \pi_d + \eta_{dA} \pi_0 - i\epsilon_{dcb} \pi^c \delta^b_A \right) E_0^A \lambda^d,\\
\{ \sigma_d[\lambda^d], \sigma_e[\kappa^e] \} &= \sigma_c [2i \epsilon^{cab} \lambda_a \kappa_b],
\end{align} 
which is a smeared version of the Lie algebra $\isl(2, \C)$.

\section{Conclusion}
In this paper, the Hamiltonian analysis for the self-dual gauge gravity theory has been carried out. The canonical structure is identical to that of normal self-dual gravity, and thus the gauge gravity formalism offers no obvious novelty in the canonical quantisation process.

\section{Acknowledgements}
I thank John Barrett and Lee Smolin for discussions. This work was funded by the Leverhulme trust.

\end{document}